# Etude de la composition chimique et du potentiel pharmacologique associé de *Phyllanthus amarus Schum et Thonn. (1827)*

## « Grenn anba fèy »

Mélissa MATOU[1], Sylvie BERCION[1,2], Patrick MERCIRIS[1], Nicole LAURENT MEYSSONIER[1], Déborah FERNAND[1], Thérèse MARIANNE-PEPIN[1]

[1] UMR Qualitrop. Université des Antilles (Campus de Fouillole), Guadeloupe.
[2] APLAMEDAROM

## INTRODUCTION

En 2013, 43 plantes issus des Antilles ont fait leur entrée à la pharmacopée française dont 15 proposé par la Guadeloupe par l'association APLAMEDAROM sur la base du recueil de données bibliographiques. Cependant, malgré leur inscription sur la liste A, les professionnels de santé restent encore sceptiques et déplorent un manque de données sur la composition des spécimens domiens et craignent les interactions médicamenteuses et les problèmes de dosage (Genthon et al., 2014)

*Phyllanthus amarus* Schum & Thonn. (1827) est très connu dans le monde notamment en Inde où la plante est très utilisée et commercialisée pour ses propriétés contre les pathologies liées au foie, aux problèmes digestifs et le diabète. Plus connue, aux Antilles Françaises et notamment en Guadeloupe, sous le nom créole de Grenn anba fèy, elle est aussi couramment consommée entière sous forme d'infusion ou de décoction aqueuse afin de lutter contre le diabète, l'hypentension et le cholestérol.

Beaucoup utilisée et étudiée en Inde et en Afrique, on retrouve dans la littérature, un certain nombre de données sur sa composition chimique. Il est notamment reporté que la plante comporte des métabolites secondaires qui lui confèrent ses activités biologiques. Dans cette catégorie, l'on retrouve dans les extraits aqueux, des huiles essentielles et terpènes, des alcaloïdes et des polyphénols auxquels nous nous intéresserons plus particulièrement. Parmi eux, quatre familles de molécules bioactives majoritaires sont identifiées dans les extraits : des acides phénoliques, des flavonoïdes, des lignanes et des tanins.

Nous nous intéressons à trois activités biologiques de la plante, représentatives des pathologies couramment retrouvées aux Antilles : l'activité antioxydante, anti-inflammatoire et antidiabétiques.

Le stress oxydatif est un état physiologique résultant d'un déséquilibre entre le système de défense antioxydante et les espèces oxydantes. Il est à l'origine de la survenue ou de l'aggravation de nombreuses pathologies telles que les maladies neurodégénératives (Alzheimer, Pakinson), les maladies cardiovasculaires (hypertension) ou encore les cancers. La littérature relate différents mécanismes d'action notamment par l'étude de l'action des extraits aqueux sur les enzymes antioxydantes, le potentiel de piégeage de radicaux libres, l'inhibition de la peroxydation lipidique et le potentiel de réduction du fer. Les polyphénols ont la particularité d'être tous antioxydants (Guha et al., 2010 ; Patel et al., 2011) et sont donc soupçonnés d'être à l'origine de cette activité.

L'activité anti-inflammatoire a été étudiée dans la littérature sur les extraits aqueux notamment par des tests sur animaux. Les maladies inflammatoires sont aussi très répandues aux Antilles. Nous avons l'exemple de la drépanocytose, qui est due à une mutation génétique à l'origine d'une perte de la capacité des érythrocytes à se déformer, dont la prévalence est forte dans les populations Antillaises. Les extraits de *P. amarus* ont justement montrés une action stabilisatrice sur la membrane des érythrocytes via la modification de l'influx calcique ; ce qui est étroitement lié à la déformabilité et au volume des érythrocytes intervenant dans la maladie de la drépanocytose. Les lignanes et les flavonoïdes seraient responsables de cette activité (Atul R. Chopade et al., 2012).

Une étude de Iranloye, en 2011, a montré une diminution de 70% de l'œdème de la patte avec 200 mg/kg d'extrait qui était supérieur au résultat obtenu pour le médicament de référence. Cette activité est attribuée à la présence des lignanes.

Le diabète est une maladie qui touche beaucoup d'antillais. Elle résulte du maintien de l'augmentation du glucose sanguin provoqué par une incapacité de l'organisme à diminuer ce taux. Il existe deux types de diabète : le diabète insulinodépendant (DT1) et le diabète non-insulinodépendant (DT2). Les extraits de *P. amarus* ont montré une activité antidiabétique par la diminution du taux de glucose sanguin (Mbagwu et al., 2011) et l'inhibition des enzymes digestives qui permet donc de ralentir la digestion des sucres (Patel et al., 2011). Ces tests sur animaux ont permis de démontrer une action positive contre le diabète de type 2 (DT2).
Ces activités sont essentiellement attribuées aux composés terpéniques en particulier les diterpènes.

La composition chimique des plantes varie selon leur lieu de récolte et peut donc avoir un impact sur les mécanismes d'action d'où la nécessité d'étudier les spécimens issus des Antilles. Notre étude vise à identifier et/ou caractériser les composés responsables des activités observées par la population notamment ceux responsables des activités anti-oxydantes, antidiabétiques et anti-inflammatoires. Nous avons cherché à caractériser les spécificités d'extraits aqueux obtenus à partir des spécimens récoltés en Guadeloupe. Avec l'approche de déréplication, nous avons cherché à mettre en évidence les différences entre les spécimens étudiés avec ceux dont l'étude a été reporté dans la littérature afin de mieux comprendre les trois activités biologiques sélectionnées.

## MATERIELS ET METHODES

### ♦ Matériel

*Echantillonnage*

Les parties aériennes fraîches de *P. amarus* ont été récoltés au Gosier (Guadeloupe) au cours du mois de Mars 2016 et préparés selon deux modes de préparation communément utilisés par la population.

L'identification de la plante a été réalisée avec l'aide de Monsieur Alain ROUSTEAU, MCF HDR de l'Université des Antilles.

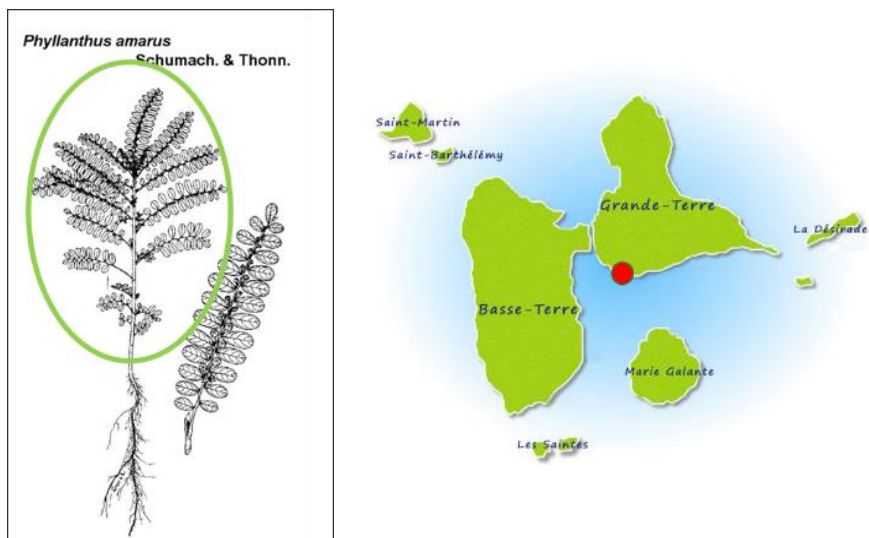

**Figure 1: Schéma et zone de récolte de Phyllanthus amarus**

Les extraits aqueux sont préparés à une concentration égale à 30 g/L selon ces deux modes de préparations et ont étés réalisés en triplicata.

## Méthodes

*Préparation des infusés aqueux*

Cette étape consiste à préparer une infusion aqueuse des parties aériennes fraîches à l'aide d'eau filtrée avec un dispositif millipore (figure 2)

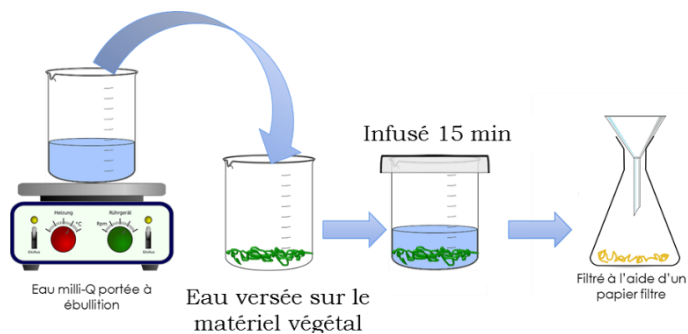

**Figure 2: Etapes de préparation d'une infusion aqueuse**

*Préparation des décoctés aqueux*

Cette étape consiste à préparer une décoction aqueuse des parties aériennes fraîches a l'aide d'eau filtrée avec un dispositif millipore (figure 3)

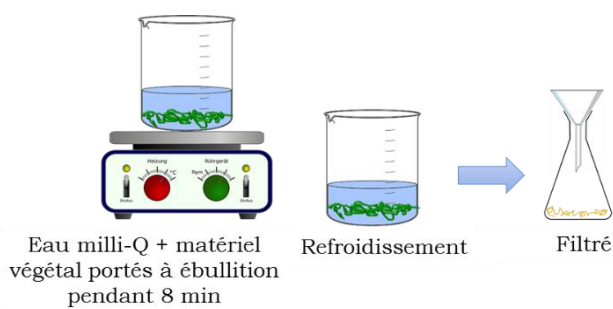

**Figure 3: Etapes de préparation d'une décoction aqueuse**

Les extraits aqueux obtenus sont ensuite évaporés à l'évaporateur rotatif (T = 40°C), afin d'obtenir des extraits secs, puis placés au dessiccateur toute la nuit avant d'être pesés.

Les extraits secs sont, ensuite, conservés au congélateur -80°C jusqu'à analyse.

---

*Analyse par Chromatographie en phase Liquide à Ultra Haute performance couplée à un spectromètre de masse (UHPLC-SM) / QUALITROP*

L'analyse des extraits bruts de *P. amarus* a été réalisée sur une chaîne UHPLC (Dionex Ultimate 3000) couplée à un spectromètre de masse ORBITRAP à haute résolution (Exactive plus) (Figure 4).

On utilise une colonne analytique ZORBAX Eclipse XDB-C18 (porosité 5µm ; longueur : 150 mm ; diamètre : 4.6 mm) et un système d'éluant constitué de d'un solvant A (eau 98% + acide formique 2%) et solvant B (acétonitrile 80% + eau 18% + acide formique 2%) distribués sous forme de gradient.

Le choix des longueurs d'ondes étudiées a été déterminé par le type de composés recherchés. Les composés polyphénoliques absorbent majoritairement entre 190 et 360 nm d'où le choix de ces longueurs d'onde de détection. Deux longueurs d'ondes ont donc été sélectionnées :

- 280 nm, permettant d'observer préférentiellement la famille des acides phénoliques, des tanins, des terpènes et des lignanes
- 360 nm où la famille des flavonoïdes absorbe préférentiellement

La détection par spectrométrie de masse a été réalisées à l'aide d'un appareil de type Exactive plus (Thermofischer), constitué d'une source d'ionisation par électrospray (ESI) et d'un analyseur Orbitrap (trappe piégeuse d'ions).

Le spectre de masse obtenu correspond à l'abondance relative de l'ensemble des ions générés en fonction de leur rapport m/z.

## RESULTATS ET DISCUSSIONS

L'analyse des profils chromatographiques permet de distinguer quatre familles de composés localisés en fonction de leur polarité (Figure 9).

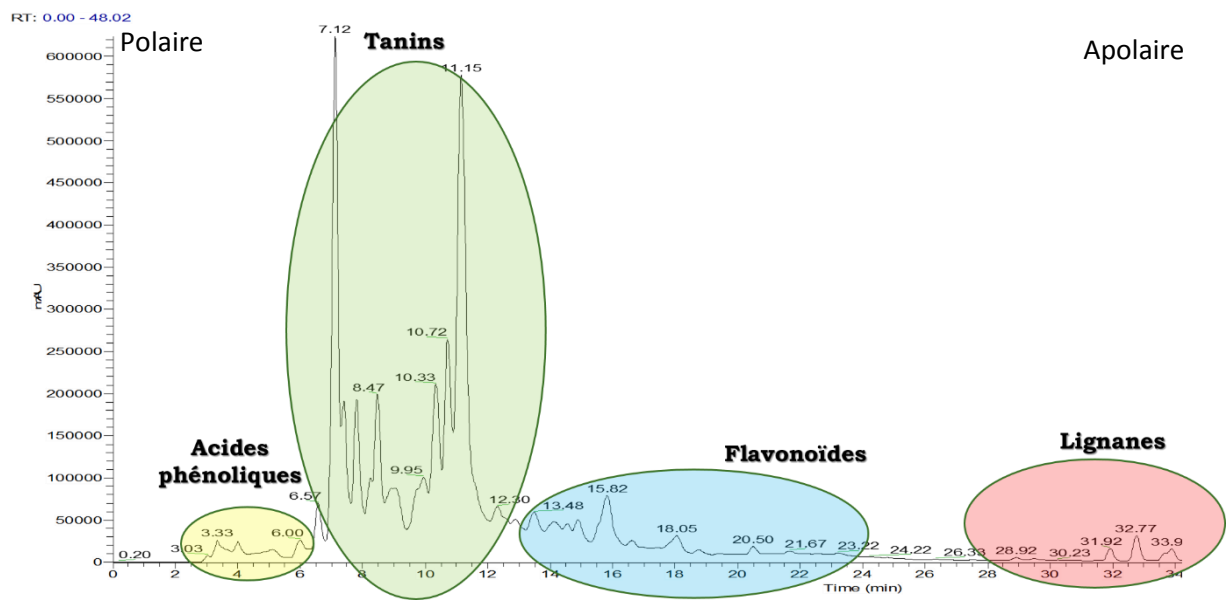

**Figure 4: Chromatogramme UV d'un infusé aqueux des parties aérienne de *Phyllanthus amarus* à 280 nm**

Les acides phénoliques se retrouvent majoritairement en début de chromatogramme. Les tanins, qui sont ensuite préférentiellement localisés entre 7 et 12 min et semblent être majoritairement dans les extraits aqueux de la plante. Puis les flavonoïdes et les lignanes se retrouvent majoritairement en fin de chromatogrammes.

L'étude des extraits par déréplication, a permis d'identifier 31 molécules dans l'infusé et/ou le décocté aqueux des parties aériennes de *Phyllanthus amarus*.

Le tableau 8 donne la liste de ces molécules avec l'ensemble des données collectées ayant permis leur identification. La liste est classée par ordre des temps de rétention et un numéro d'identification a été attribué à chaque composé. Les formules brutes et les masses molaires indiquées ont été obtenues par la littérature.

L'analyse des profils chromatographiques des extraits aqueux des parties aériennes fraîches de la plante montre quelques différences différence.

**Tableau 1: caractéristiques chromatographiques des molécules identifiées dans les extraits aqueux de *Phyllanthus amarus***

| Caractéristique chromatographique | | Analyse UHPLC - SM | | | | Identification du composé | | | |
|---|---|---|---|---|---|---|---|---|---|
| N° de composé | tR (min) | Ion observé m/z ESI (-) | Ion observé m/z ESI (+) | Ion calculé | Ion fils obtenus en SM2 [M/Z (abondance relative)] | Métabolite identifié | Formule brute | Masse moléculaire (Da) | Ref |
| 2 | 3,97 | 191,0192 | | 191,01880 | | Acide mucique lactone | $C_6H_8O_7$ | 192,02663 | |
| 3 | 5,15 | 331,06784 | | 331,06653 | 211,02490 (30); 169,01381 (100) | Monogalloyl-hexoside | $C_{13}H_{16}O_{10}$ | 332,07435 | |
| 4 | 5,91 | 169,01353 | | 169,01370 | | Acide gallique | $C_7H_6O_5$ | 170,02153 | |
| 13 | 10,37 | 291,01474 | 293,03073 | 291,01450 | 247,02502 (100); 219,02979 (10); 191,03430 (31); 175,03943 (5) | Acide carboxylique de Brévifoline | $C_{13}H_8O_8$ | 292,02233 | |
| 14 | 10,71 | 633,07379 | | 633,07280 | 481,06235 (2); 463,05255 (30); 300,99930 (100); 275,02008 (18) | Corilagine | $C_{27}H_{22}O_{18}$ | 634,08062 | |
| 15 | 11,20 | 951,07697 | | 951,07398 | 933,06750 (100); 300,99982 (72); 169,01366 (30) | Geraniine | $C_{41}H_{28}O_{27}$ | 952,08181 | |
| 16 | 11,35 | 969,08563 | | 969,08455 | | Acide amariinique | $C_{41}H_{30}O_{28}$ | 970,09237 | S. Kumar et al, 2015; A D. Sousa et al, 2016 |
| 19 | 12,27 | 925,10193 | | 925,09783 | | Phyllanthusiine C | $C_{40}H_{30}O_{26}$ | 926,10565 | |
| 20 | 12,59 | 463,05276 | | 463,05220 | 300,99942 (100); 283,99545 (1); 257,00964 (1); 169,01360 (6) | Ellagic acid -O-hexoside | | 464,06003 | |
| 21 | 13,44 | 305,03082 | | 305,03030 | | Methylbrevifolincarboxylate | | 306,03813 | |
| 24 | 15,62 | 609,14722 | | 609,14557 | | Quercétine - 3 -O rutinoside (rutine) | $C_{27}H_{30}O_{16}$ | 610,15339 | |
| 25 | 15,84 | 300,99942 | | 300,99980 | 283,99680 (15); 257,00955 (5) | Acide ellagique | $C_{14}H_6O_8$ | 302,00763 | |
| 26 | 18,05 | 923,08173 | | 923,07600 | | Phyllanthusiine U | $C_{40}H_{28}O_{26}$ | 924,08383 | |
| 28 | 23,22 | 206,97241 (100) 363,01886 (92) | 365,03513 | 365,03395 | | Niruriflavone | $C_{16}H_{12}O_8S$ | 364,02613 | |
| 29 | 31,92 | | 439,21201 | 439,17327 | | Virgatusine | $C_{23}H_{28}O_7$ | 416,18350 | |
| 30 | 32,80 | | 441,22736 | 441,22531 | | Phyllanthine | $C_{24}H_{34}O_6$ | 418,23554 | |
| 31 | 33,93 | | 455,20709 | 455,20458 | | Niranthine | $C_{24}H_{32}O_7$ | 432,21481 | |

(composés identifiés par co-injection de standard, RMN ou comparaison des masses obtenues avec la littérature)

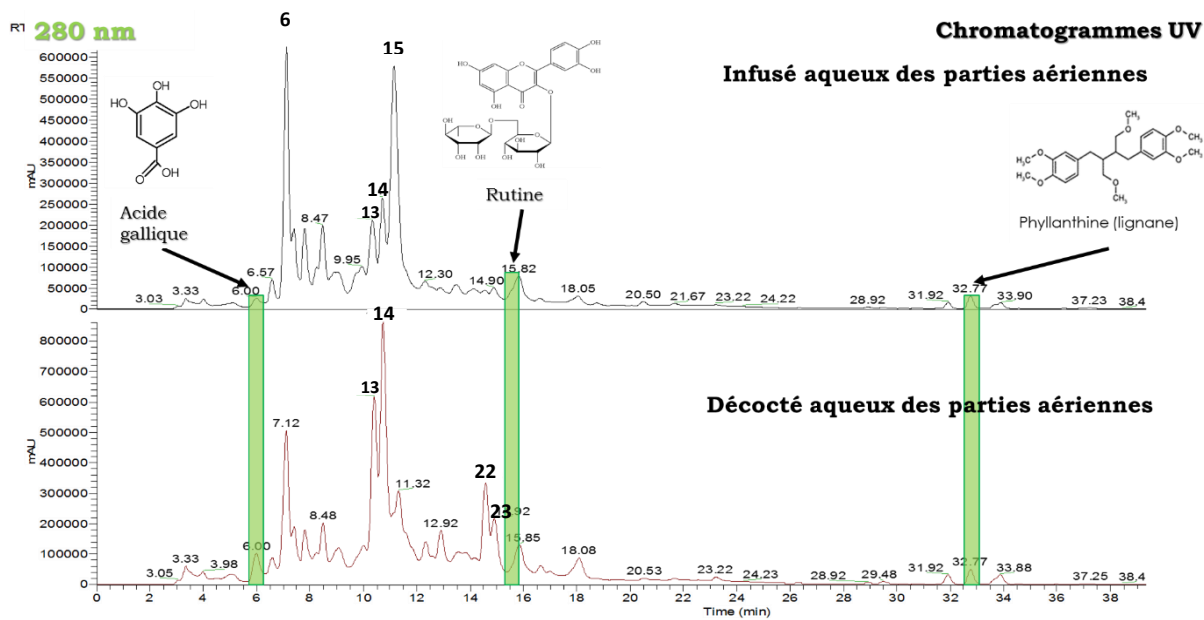

**Figure 5: Chromatogrammes UV de l'infusé et du décocté aqueux des parties aériennes fraîches de *Phyllanthus amarus* à 280 nm**

Concernant les deux modes de préparation utilisés par la population, nous pouvons constater que d'un point de vue qualitatif, les chromatogrammes sont très semblables. Cependant, la différence d'intensité laisse à penser que le mode de préparation pourrait influencer la quantité de composés extraits mais une analyse quantitative, en prenant en compte les aires des pics, est nécessaire pour l'affirmer.

Nous constatons que les composés majoritaires de l'infusé sont le **composé 6** et la **géraniine (15)**, alors que dans le décocté se sont la **corilagine (14)** et l'**acide carboxylique de Brévifoline (13)** qui se retrouvent majoritairement dans les extraits. La **géraniine (15)** est un ellagitanin hydrosoluble aux propriétés antioxydante, antibactérienne, antivirale, anticancéreuse, immunomodulatrice, hépatoprotectrice, anti-hypertensive antidiabétique et bien d'autres. Cette molécule, facilement hydrolysable en présence d'eau chaude, s'hydrolyse en deux autres composés : l'acide gallique et l'acide ellagique ainsi qu'un glucose (Hong Sheng Cheng, So Ha Ton, & Khalid Abdul Kadir, 2016). Ce phénomène pourrait expliquer la perte de géraniine dans les extraits de décocté aqueux. La **corilagine (14)** (Figure 13) est aussi un ellagitanin aux propriétés anti-hypertensive, anti-tumorale, antioxydante, anti-inflammatoire et antivirale (Weiyan, Lei, & Kun, 2014).

Peu d'études ont été réalisées sur **l'acide carboxylique de Brévifoline (13)**, cependant, la littérature relate une activité cytotoxique contre des cellules cancéreuses (Ihn Rhan Lee & Mi Young Yang , 1994) ainsi qu'une activité anti-oxydante (Sónia A.O. Santos, et al., 2013).

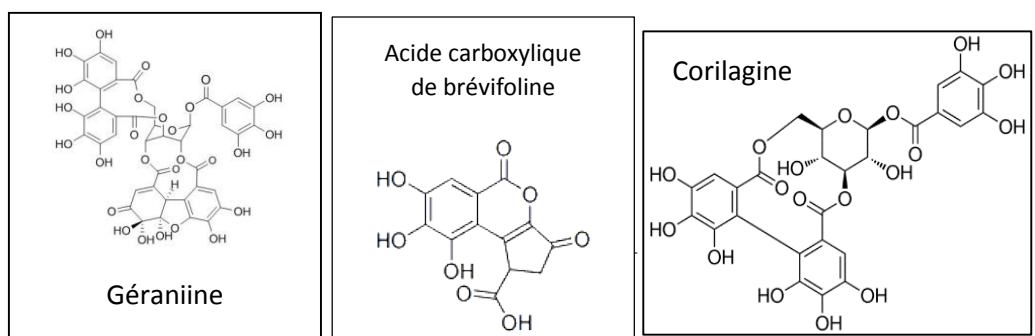

**Figure 6: Formules développées des composés 13, 14 et 15**

De plus, deux molécules (**22** et **23**) sont retrouvées dans les extraits de décocté aqueux des parties aérienne fraîches mais en très faible intensité dans les extraits d'infusé aqueux. Ces deux composés ne sont pas répertoriés dans la littérature pour l'espèce *P. amarus* et restent encore à être caractérisés.

**CONCLUSION ET PERSPECTIVES :**

D'un point de vue qualitatif, il semblerait que la composition chimique des deux types d'extraits soit semblable, cependant, une analyse quantitative s'avère nécessaire afin d'étudier l'influence du mode de préparation sur les composés extraits.

Nous avons donc pu identifier pour les parties aériennes :
- Formellement 3 molécules par comparaison avec des standards disponibles, l'acide gallique (acide phénolique), la rutine (flavonoïde) et la phyllanthine (lignane).
- 13 molécules suspectées d'être des composés déjà identifiés dans la littérature pour cette espèce par la méthode de déréplication : des **acides phénoliques** (acide mucique lactone, monogalloyl-hexoside, acide ellagique), des **tanins** (syringine, corilagine, acide carboxylique de brevifoline, geraniine, acide aramiinique et phyllanthusine C), un **flavonoïde** (niruriflavone) ainsi que des **lignanes** (virgatusine, niranthine)
- 15 molécules non répertoriées dans la littérature

La présence de ces familles de composés permet déjà de mieux comprendre l'origine des propriétés observées par la population. La phyllantine, caractéristique de l'espèce, est une lignane très active ayant une forte activité antioxydante, hépatoprotectrice et anticancéreuse (Patel et al., 2011). La rutine est connue pour ses propriétés anti-oxydante et anti-hypertensive (Patel et al., 2011). L'acide gallique se distingue par des propriétés anti-oxydante, anti-inflammatoire, antidiabétique, antivirale et anticancéreuse et a l'avantage d'être un composé biodisponible.

La présence de lignanes identifiées pour certaines et à caractériser pour d'autres pourrait expliquer les propriétés anti-inflammatoires de la plante. En ce qui concerne les propriétés antidiabétiques attribuées aux composés terpéniques, la présence de composés polyphénoliques au potentiel antidiabétique tels que l'acide carboxylique et la syringine, nous permettent de soupçonner leur rôle sur le diabète.

Dans le but de mieux comprendre ces mécanismes, nous étudierons les propriétés antioxydantes des extraits de la plante par des tests in vitro et in vivo tels que DPPH, FRAP et ORAC afin de tester leur capacité à piéger les radicaux libres (peroxyles ou hydroxyles) et leur capacité à réduire le fer.

Dans le but d'étudier leur implication dans l'activité anti-inflammatoire, nous allons tester leur capacité à inhiber la sécrétion de cytokines par des macrophages activés par un inducteur de l'inflammation (LPS). Nous testerons aussi la capacité des composés à inhiber les produits terminaux de la glycation (AGE) et à inhiber l'action de l'α-glucosidase, afin d'étudier leur action antidiabétique.

Et enfin, l'étude de l'impact de la digestion, nous permettra donc de vérifier le devenir de ces composés après les avoir ingéré. Pour cela, nous réaliserons des tests sur la dégradation des composés par l'intervention d'enzymes salivaire et gastro-intestinale.

# Références

APLAMEDAROM. (s.d.). *Synthèse bibliographique. Graines en bas feuilles, Phyllantus amarus Schum. Et Thonn. = P. niruri var. amarus (Schum. et Thonn.) Leandri.*

Atul R. Chopade, Prakash M. Somade, & Fahim J. Sayyad. (2012). Membrane stabilizing activity and protein denaturation: a possible Mechanism of action for the anti-inflammatory activity of Phyllanthus amarus. *Journal of Kirshna Institute of Medicinal Sciences University*, Vol.1 (1) 67-72.

D. B. James, O. A. Owolabi, N. Elebo, S. Hassan, & L. Odemene. (2009). Glucose tolerance test and some biochemical effect of Phyllanthus amarus aquoeus extacts on normaglycemic albino rats. *African Journal of Biotechnology Vol. 8 (8)*, 1637-1642.

Genthon, L., Paul, J.-l., Nasso, Y., Gustave, M., & Bercion, S. (2014). Quel potentiel de développement en Guadeloupe de l'utilisation de plantes médicinales inscrites à la phramacopée française en août 2013?

Hasenah Ali, P.J. Houghton, & Amala Soumyanath. (2006). a-Amylase inhibitory activity of some Malaysian plants used to treat diabetes; with particular reference to Phyllanthus amarus. *Journal of Ethnopharmacology 107*, 449–455.

Hong Sheng Cheng, So Ha Ton, & Khalid Abdul Kadir. (2016). Ellagitannin geraniin: a review of the natural sources, biosynthesis, pharmacokinetics and biological effects. *Phytochem Rev*.

Ihn Rhan Lee, & Mi Young Yang. (1994). Phenolic Compounds from Duchesnea chrysantha and their Cytotoxic Activities in Human Cancer Cell. *Arch. Pharm. Res. VoL 17, No. 6*, 476-479.

Lantto, T.A, H.J.D. Dorman, A. N. Shikov, O. N. Pozharitskaya, V. G. Makarov, . . . A. Raasmaja. (2009). Chemical composition, antioxidative activity and cell viability effects of a Siberian pine (Pinus sibirica Du Tour) estract. *Food Chemistry 112 (4)*, 936-943.

Patel, JP, Tripathi, P, Sharma, V, Chauhan, NS, Di. (2011). Phyllanthus amarus : Ethnomedical uses, phytochemistry and pharmacology: a review. *Journal of Ethnopharmacology*, 286-313.

S Tawata, S. T. (1996). Syntheses and biological activities of dihydro-5,6-dehydrokawain derivatives. . *Biosci. Biotechnol. Biochem.60*, 1643-1645.

Sónia A.O. Santos, Juan J. Villaverde, Andreia F. Sousa, Jorge F.J. Coelho, Carlos P. Netoa, & Armando J.D. Silvestre. (2013). Phenolic composition and antioxidant activity of industrial cork by-products. *Industrial Crops and Products (47)*, 262–269.

Soumya Maity, Suchandra Chatterjee, Prasad Shekhar Variyar, Arun Sharma, Soumyakanti Adhikari, & Santasree Mazumder. (2013). Evaluation of Antioxidant Activity and Characterization of Phenolic constituents of phyllanthus amarus roots. *Journal of Agricultural and food chemistry*, 3443-3450.

Weiyan, Q., Lei, H., & Kun, G. (2014). Chemical Constituents of the Plants from the Genus Phyllanthus. *CHEMISTRY & BIODIVERSITY*, 364-395.